\documentclass{aa}
\usepackage{graphicx}
\usepackage{epsfig}
\def\ltsima{$\; \buildrel < \over \sim \;$}
\def\gtsima{$\; \buildrel > \over \sim \;$}
\def\simlt{\lower.5ex\hbox{\ltsima}}
\def\simgt{\lower.5ex\hbox{\gtsima}}

\begin{document}

   \title{XMM-Newton observation of the bright Seyfert 1 galaxy, MCG+8-11-11}

   \author{G. Matt\inst{1},  S. Bianchi\inst{1,2}, A. De Rosa\inst{3}, P. Grandi\inst{4},
G.C. Perola\inst{1} 
          }

   \offprints{G. Matt}

   \institute{Dipartimento di Fisica, Universit\`a degli Studi Roma Tre,
via della Vasca Navale 84, I-00146 Roma, Italy
\and 
XMM-$Newton$ Science Operation Center, ESAC/ESA, Apartado 50727, E--28080 Madrid, Spain 
\and
IASF - Sezione di Roma, INAF, via del Fosso del Cavaliere 100, 00113 Roma, Italy
\and 
IASF - Sezione di Bologna, INAF, Via Gobetti 101, I-40129, Bologna, Italy
}

   \date{Received ; Accepted }

   \abstract{ We report on the 
XMM-$Newton$ observation of the bright Seyfert 1 galaxy, MCG+8-11-11.
Data from the EPIC/p-n camera, the Reflection Gratings Spectrometers
(RGS) and the Optical Monitor (OM) have been analyzed. 
The p-n spectrum is well fitted by a power law, a spectrally unresolved  Fe K$\alpha$ line,
a Compton reflection component (whose large value, when compared to the iron line
equivalent width, suggests iron underabundance), and absorption by warm material.
Absorption lines are apparent in the RGS spectra, but their
identification is uncertain and would require large matter velocities.
The UV fluxes measured by the OM are well
above the extrapolation of the X-ray spectrum, indicating the presence of a UV bump.
   \keywords{Galaxies: active -- X-rays: galaxies -- Seyferts:
individual: MCG+8-11-11
               }}

\authorrunning{G. Matt et al.}
\titlerunning{XMM-$Newton$ observation of the bright Seyfert 1 galaxy, MCG+8-11-11}

   \maketitle
%

\section{Introduction}

One of the main surprise from $Chandra$ and especially XMM-$Newton$ observations of 
Seyfert galaxies is that the iron K$\alpha$ line, while confirmed to be almost
ubiquitous, is often (but not always) spectrally unresolved, and therefore 
unlikely to originate in the
innermost regions of the accretion disc. While relativistically broadened lines have 
sometimes been confirmed (e.g. in MCG-6-30-15, Wilms et al. 2001 and Fabian et al. 2002; 
NGC~3516, Turner et al. 2002; MCG-5-23-16, Balestra et al. 2004), in many objects
only the unresolved component is visible, the upper limits to the relativistic component
being quite tight (e.g. Bianchi et al. 2004). The reason for a different behaviour
from source to source is at present unclear. Among the proposed explanations are disc
truncation and significant ionization of the matter.

MGC+8-11-11 ($z$=0.0205) is one of the brightest AGN in the X-ray band. 
It has been observed by all major X-ray satellites, with the notable exception
of $Chandra$. ASCA (Grandi et al. 1998) and BeppoSAX (Perola et al. 2000) found that
the spectrum is well fitted by a pretty standard model composed of a power law,
a  warm absorber, a Compton reflection component, and an iron K$\alpha$ line. The moderate
energy resolution and/or sensitivity of those instruments left undecided whether the
line is orginated in the innermost regions of the accretion disc (and thence relativistically
broadened) or in distant matter (and thence unresolved). One of the main goals of 
the XMM-$Newton$ observation of this source is to solve this issue. 

\section{Observation and data reduction}

MCG+8-11-11 was observed by XMM-$Newton$ on 10 April 2004 [\textsc{Obsid} 0201930201] 
with the EPIC p-n (Str\"uder et
al. 2001) and MOS (Turner et al. 2001) cameras, 
both in small window/medium filter mode, with the RGS and the OM. 
Data were reduced with SAS 6.0.0 adopting standard procedures, while screening for 
intervals of flaring particle
background in the EPIC data was done consistently with the
choice of the extraction radii, in order to maximise the
signal-to-noise ratio, similarly to what described by Piconcelli et al. (2004). 
As a result of this procedure, the
effective time and the source extraction radius in the p-n are 37.7 ks and $40$
arcsec. Patterns 0 to 4 were included in the p-n spectrum, whose count rate (14.1 counts/s) 
is lower than the maximum for 1 per cent
pile-up (see Table 3 of Ehle et al. 2005). Effective time in the RGS is 38.2 ks.

We do not make use of MOS data in this paper for two reasons: first, the MOS are slightly
piled-up, despite the small window mode adopted; second, just because of the small window
it is very difficult to extract the background in the field-of-view, and the use of blank
fields is then necessary, which may oversubtract the spectrum at low energies due to
the high Galactic column to the source.

p-n spectra were rebinned to oversample the energy resolution by a factor about 3 and,
if necessary, further rebinned until at least 25 counts/bin are obtained. For the gratings spectra,
only the latter criterium has been adopted. 

Six exposures with the Optical Monitor were also available, three with the UVW1 (291 nm)
and three with the UVW2 (212 nm) filters. 

In the following, all errors will refer to 90\% confidence level for one interesting parameter
($\Delta\chi^2$=2.71).


\section{Results}

\subsection{Temporal analysis}

\begin{figure}
\epsfig{figure=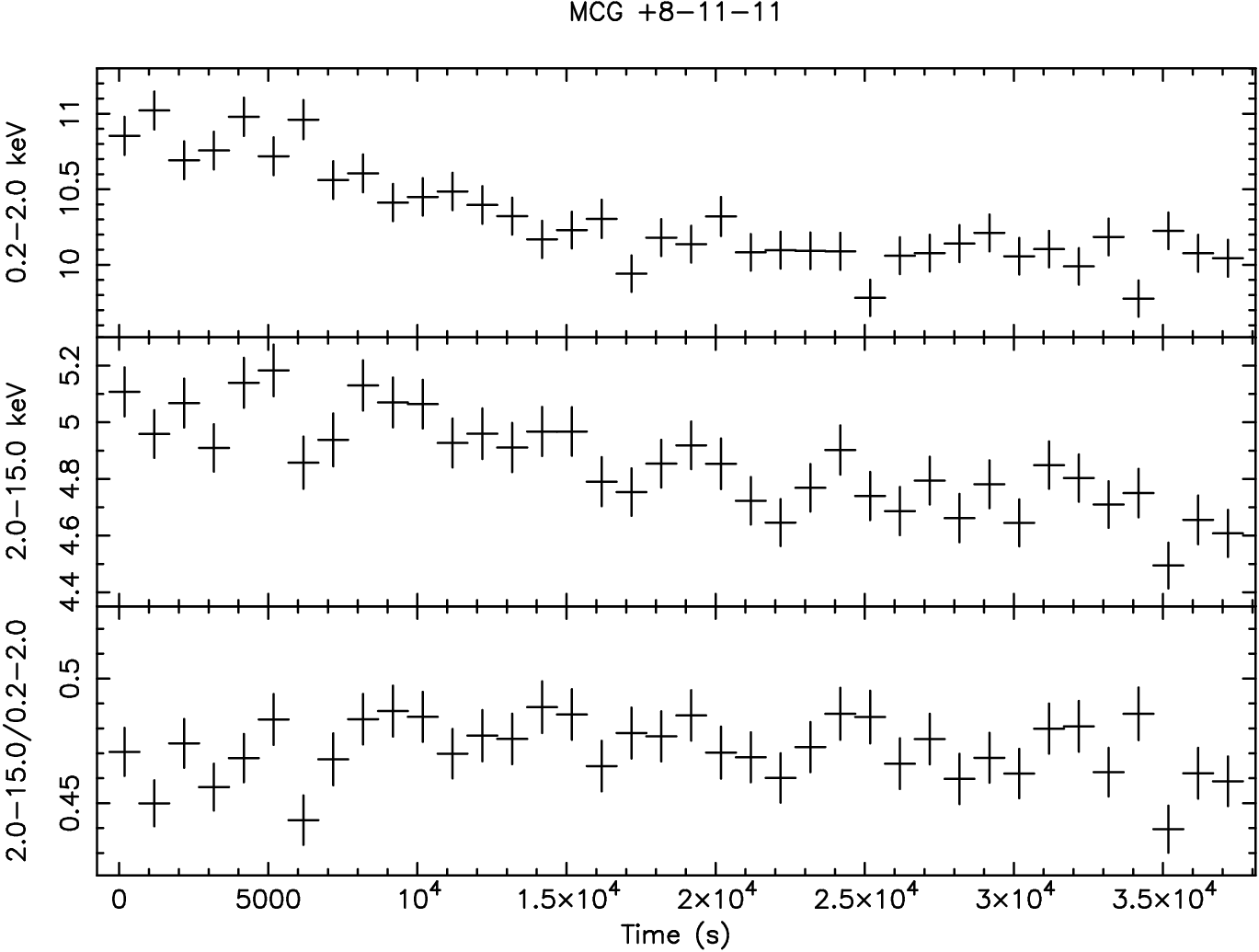,height=8.0cm,width=7.0cm,angle=-90}
\caption{The lightcurves in the 0.2-2 keV (upper panel) and 2-15 keV band (middle panel), 
and of the 2-15/0.2-2 keV hardness ratio.}
\label{hardratio}
\end{figure}

During the observation, the source flux varied by about 10\%, with no
significant spectral variations (see Fig.~\ref{hardratio}). We then  divided 
the observation into five equally spaced intervals. After having verified that
the spectral parameters in the five intervals remain constant within the errors, we searched
for the presence of absorption or emission transient features, as those recently claimed to
be present in the spectra of many Seyfert galaxies
(e.g. Pech\'a\v{c}ek et al. 2005 and references therein; Matt et al. 2005; Dadina
et al. 2005). No significant feature has been detected.

\subsection{Spectral analysis}

\begin{figure}
\epsfig{figure=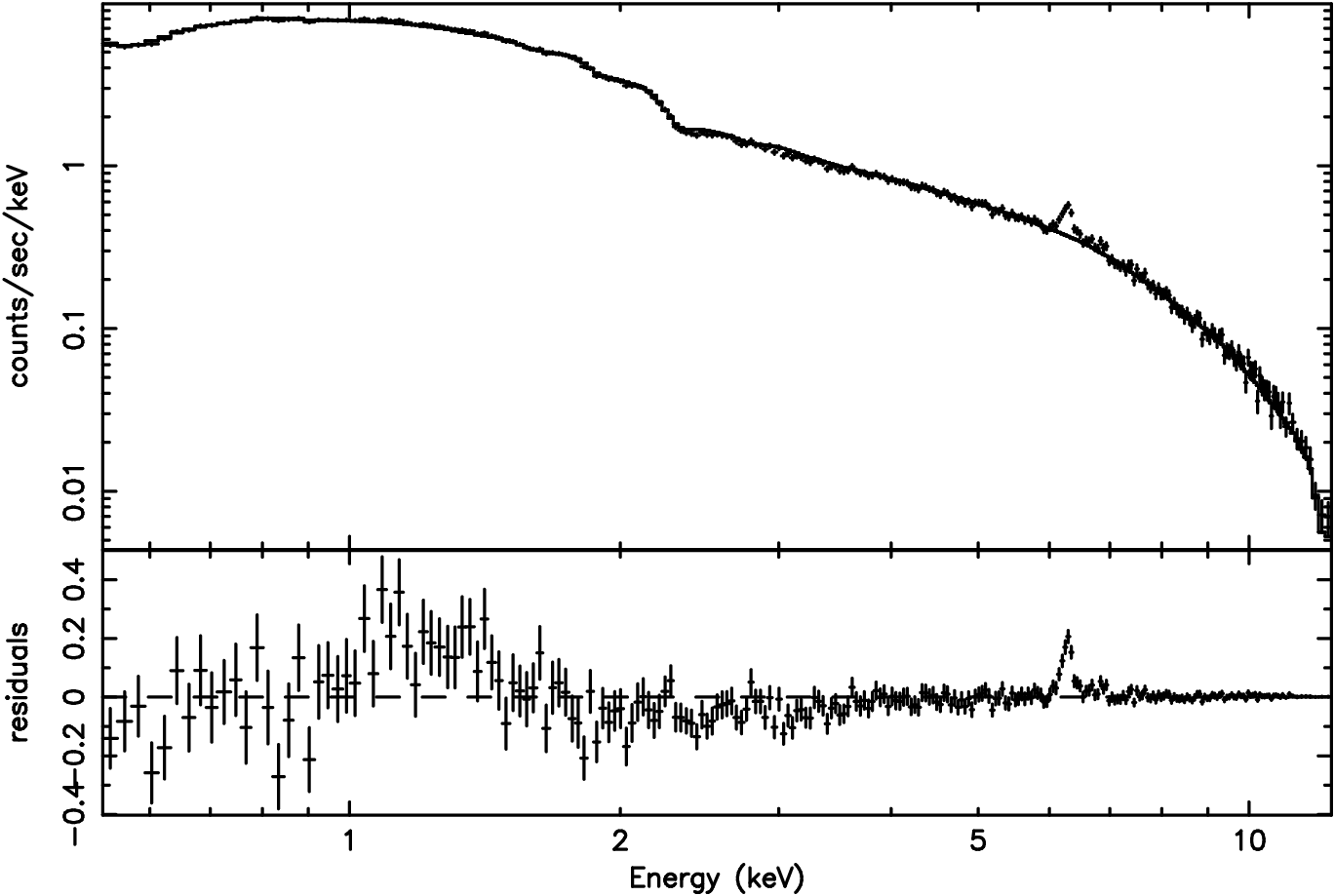,height=8.0cm,width=6.0cm,angle=-90}
\caption{Spectrum and best fit model (upper panel) and residuals (lower panel), when 
fitted with a simple absorbed power law.}
\label{badfit}
\end{figure}

\begin{figure}
\epsfig{figure=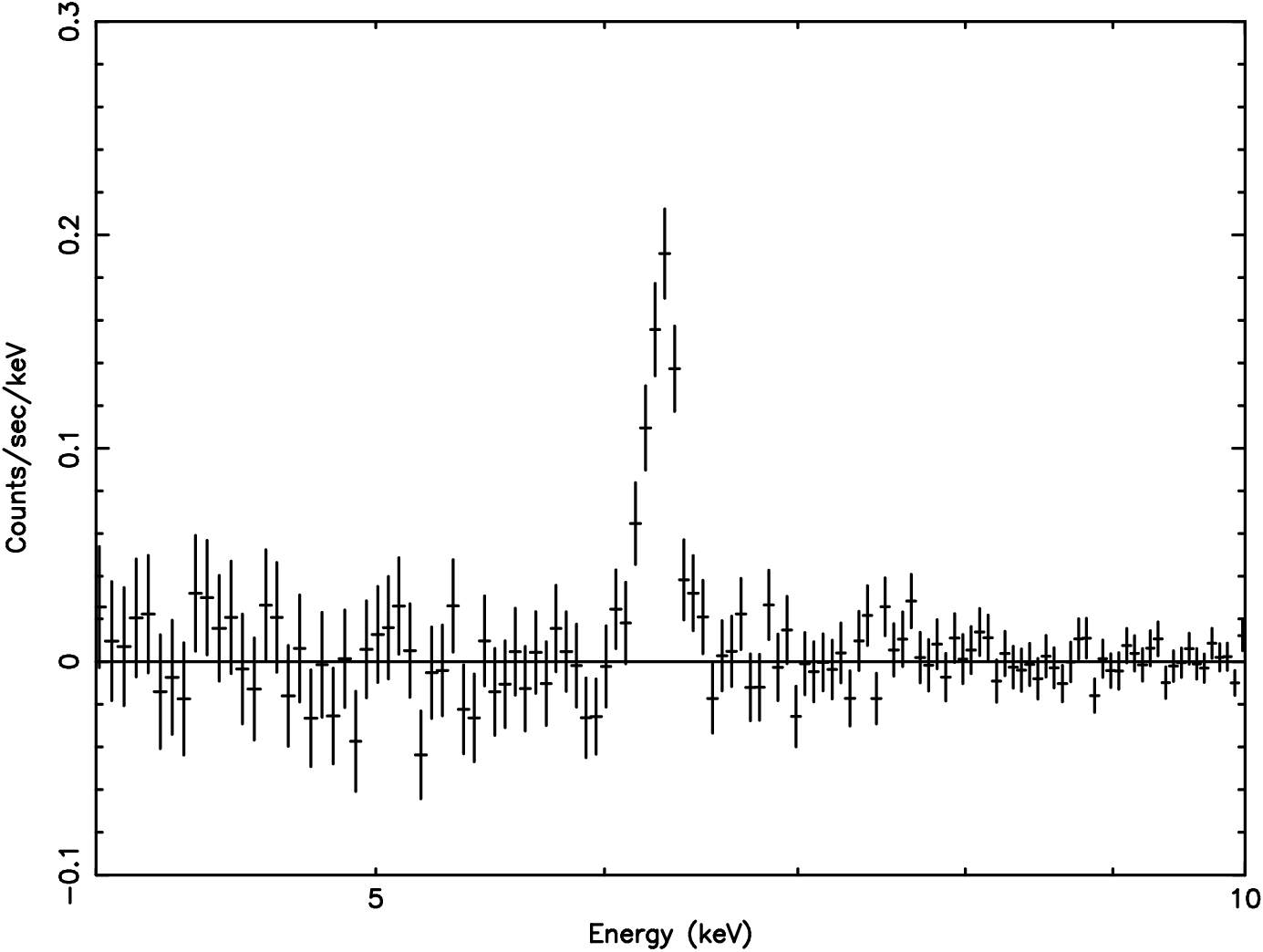,height=8.0cm,width=6.0cm,angle=-90}
\caption{Residuals around the iron line energy band. The plot
has been obtained by first fitting the spectrum with the model described
in Table~\ref{fit}, and then removing the iron line, a technique 
used for illustration purposes only.}
\label{mcg8_line}
\end{figure}

The lack of any significant spectral variability allowed us to use, for spectral
purposes, the whole observation. We started fitting the p-n instrument only, in the
0.5-12 keV energy range.
The spectrum and residuals when fitting with a simple absorbed power law are
shown in Fig.~\ref{badfit}. The fit is clearly unacceptable, with a 
$\chi^2$/d.o.f. of 697.1/260. 
An iron K$\alpha$ emission line is the most obvious feature present
in the residuals; an iron K edge and whiggles in the softer part of the spectrum are also
apparent. Therefore, and following Perola et al. (2000), we fitted
the spectrum with a model composed by an absorbed power law, a Compton reflection
component (model {\sc pexrav} in XSPEC, with the inclination angle kept fixed to
30$^{\circ}$), a gaussian iron K$\alpha$ line, and warm absorption
(model {\sc absori}, with the temperature of the material kept fixed to 10$^6$ K). Metal
abundances have been kept fixed to cosmic ones, according to Anders \& Grevesse (1989).
All these components are required by the data. The 
addition of the iron line, the reflection component and the warm absorber in turn
gives values of $\chi^2$/d.o.f of 393.4/257, 278.4/256 and 256.8/254. 
The final fit is perfectly acceptable. The iron line
is unresolved ($\sigma <$60 eV; see also Fig.~\ref{mcg8_line}
and therefore we decided, in the following fits, to fix its width
to 1 eV for the sake of simplicity. 

Inspection of the residuals showed a deficit of counts around 0.7-0.9 keV.
This suggests that a single zone warm absorption model is insufficient to describe the data.
We therefore added an edge at 0.74 keV, corresponding to He--like Oxygen.  The addition of
this component leads to $\chi^2$/d.o.f=241.9 /254.

We then added the iron K$\beta$ line, the nickel
K$\alpha$ line, and the iron K$\alpha$ Compton Shoulder (CS hereinafter). All three features
are expected to go along with the iron K$\alpha$ line.  The first two
lines were modelled with unresolved ($\sigma$=1 eV) gaussians with energies fixed at 7.06 and
7.48 keV, the CS with a gaussian with centroid energy of 6.3 keV and $\sigma$=40 eV (Matt 2002).
The Compton Shoulder is marginally significant, while for the other two lines 
only upper limits are obtained.

\begin{figure}
\epsfig{figure=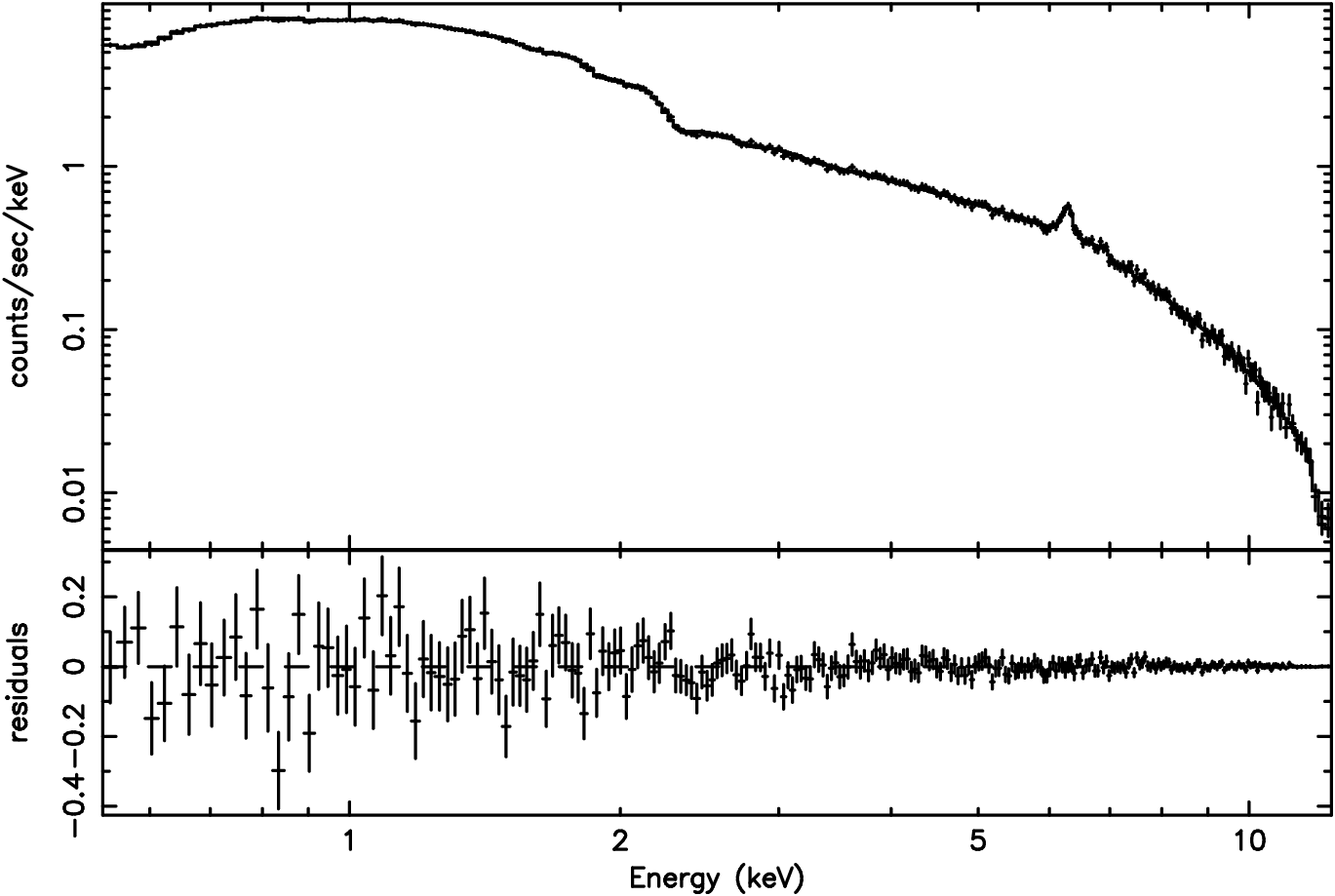,height=8.0cm,width=6.0cm,angle=-90}
\caption{Spectrum and best fit model (upper panel) and resisuals (lower panel). See
Table~\ref{fit} for the best fit parameters.}
\label{bestfit}
\end{figure}

Finally, it is worth noting that, at odds with many other Seyfert galaxies and quasars, 
no soft excess is required by the data (see Fig.~\ref{bestfit}).
 
The best fit parameters are summarized in Table~\ref{fit}, while the spectrum with the best fit
model and the residuals are shown in Fig.~\ref{bestfit}. 
The value of the column density, $\sim$1.8$\times$10$^{21}$ 
cm$^{-2}$, is consistent with the Galactic one
(Elvis et al. 1989 found a value of (2.0$\pm$0.2)$\times$10$^{21}$ 
cm$^{-2}$. Fixing the
$N_H$ to this value leads to a significantly worst, even if still perfectly acceptable fit, i.e. 
$\chi^2$/d.o.f.=255.4/252; the power law is of course steeper, i.e. 1.9, which in turn
gives a very large value of $R$, i.e., 2.6). The 2--10 keV unabsorbed
flux is  4.63$\times10^{-11}$ erg cm$^{-2}$ s$^{-1}$, 
corresponding to a luminosity in the same band
of 4.3$\times10^{43}$ erg s$^{-1}$ ($H_0$=70 km/s/Mpc). As far as we know, there is no
estimate of the Black Hole mass based on BLR reverberation mapping. Using the BLR radius
vs the 5100 \AA~ luminosity relationship (Kaspi et al. 2000) and the width of the BLR lines,
Bian \& Zhao (2003) estimated a Black hole mass of about 1.5$\times10^7$ M$_{\odot}$, implying
an Eddington luminosity of about 2$\times10^{45}$ erg s$^{-1}$. Hence, 
$L_{Bol}/L_{\rm edd}$=0.02$k$,
where $k$ is the bolometric correction, which is about 25 for the observed X-ray luminosity 
according to Marconi et al. (2004). Therefore, the source would be emitting 
at about half the Eddington
limit, which seems unlikely given the spectrum (high $\dot{m}$ sources, like NLS1 probably are, 
have usually much steeper spectra and prominent soft excesses). A more accurate Black Hole mass
estimate is needed before drawing any conclusion in this respect.

\begin{table}
\begin{center}
\begin{tabular}{|c|c|}
\hline
~ & ~ \cr
$\Gamma$ & 1.805$\pm$0.015 \cr
$N_{\rm H}$ & 1.83$^{+0.06}_{-0.03}\times$10$^{21}$ cm$^{-2}$ \cr
$R$ & 1.51$^{+0.13}_{-0.09}$ \cr
$N_{\rm H, warm}$ & 1.1$^{+0.6}_{-0.6}\times$10$^{22}$ cm$^{-2}$ \cr
$\xi$ & 460$^{+270}_{-130}$ erg cm s$^{-1}$ \cr
$\tau_{\rm edge}$ & 0.041$^{+0.018}_{-0.014}$ \cr
E(Fe K$\alpha)$ & 6.416$^{+0.014}_{-0.004}$ keV \cr
F(Fe K$\alpha)$ & 4.1$^{+0.6}_{-0.6}\times10^{-5}$  ph cm$^{-2}$ s$^{-1}$ \cr
EW(Fe K$\alpha)$ & 75$\pm$15 eV \cr
F(Fe K$\beta)$ & 4.6$^{+4.4}_{-4.5}\times10^{-6}$ ph cm$^{-2}$ s$^{-1}$ \cr
EW(Fe K$\beta)$ & $<$10 eV \cr
F(Ni K$\alpha)$ & 4.7$^{+4.5}_{-4.4}\times10^{-6}$ ph cm$^{-2}$ s$^{-1}$ \cr
EW(Ni K$\alpha)$ & $<$24 eV \cr
F(Fe K$\alpha$ CS) & 1.1$^{+0.7}_{-0.8}\times10^{-5}$ ph cm$^{-2}$ s$^{-1}$ \cr
EW(Fe K$\alpha$ CS) & 17$\pm$12 eV \cr
~ & ~ \cr
Flux (2-10 keV) & 4.63$\times10^{-11}$  erg cm$^{-2}$ s$^{-1}$ \cr
~ & ~ \cr
$\chi^2$/d.o.f. & 232.5/251 \cr
~ & ~ \cr
\hline
\end{tabular}
\caption{Parameters of the best fit model.}
\label{fit}
\end{center}
\end{table}

\subsubsection{The reflection component and the emission lines}

The iron line is unresolved, suggesting an origin in matter more distant than
the innermost regions of the accretion disc. This is similar to what found in many
other bright, local Seyfert galaxies (e.g. Bianchi et al. 2004). The clear 
presence of a reflection component (the fit without
this component gives a  $\chi^2$/d.o.f of 378.6/252) indicates that
the line emitting matter is Compton--thick (e.g. Matt et al. 2003). This is consistent
with the, admittedly loose, upper limit to the CS--to--line core ratio (about 0.4). 
Natural candidates for the line emitting region 
are the outermost part of the disc and the `torus' envisaged in Unification Models
(Antonucci 1993). In the latter hypothesis, no variability of the iron line
during the observation is expected; indeed, no variability is found. It should be noted, however, 
that if the line would follow variations of 
the continuum, 10\% variations would be expected, which are still consistent with the data.  
In the former hypothesis, the inner radius of the disc emitting region
is about 300 gravitational radii, as derived from fitting the 
iron line with a {\sc diskline} model, with inclination
angle set to 30$^{\circ}$ and outer radius to 10$^4$ gravitational radii (the fit is 
statistically as good as that with a gaussian line).


The ratio between the Fe K$\beta$ and K$\alpha$ lines is constrained to be less than
about 20\%, giving therefore no useful informations on the ionization state of iron (see
Molendi et al. 2003). A similar upper limit to the ratio holds for the Ni to Fe 
 K$\alpha$ line flux ratio, providing only a very loose constraint on the Ni-to-Fe
abundance, i.e. less than about 5. 

The iron K$\alpha$ line EW, about 90 eV (including the CS), 
is lower than expected given the large amount of the reflection
component (e.g. George \& Fabian 1991; Matt et al. 1991: note that fixing
$R$, the solid angle subtended by the reflecting matter in units of 2$\pi$,
 to 1, a still large but more reasonable value given the iron line EW,
 the fit is significantly worse, i.e. $\chi^2/d.o.f.$=283.5/252: note also that
fitting the spectrum without any warm absorption gives even higher values of $R$). 
There are two possible explanations for that. 

\begin{figure}
\epsfig{figure=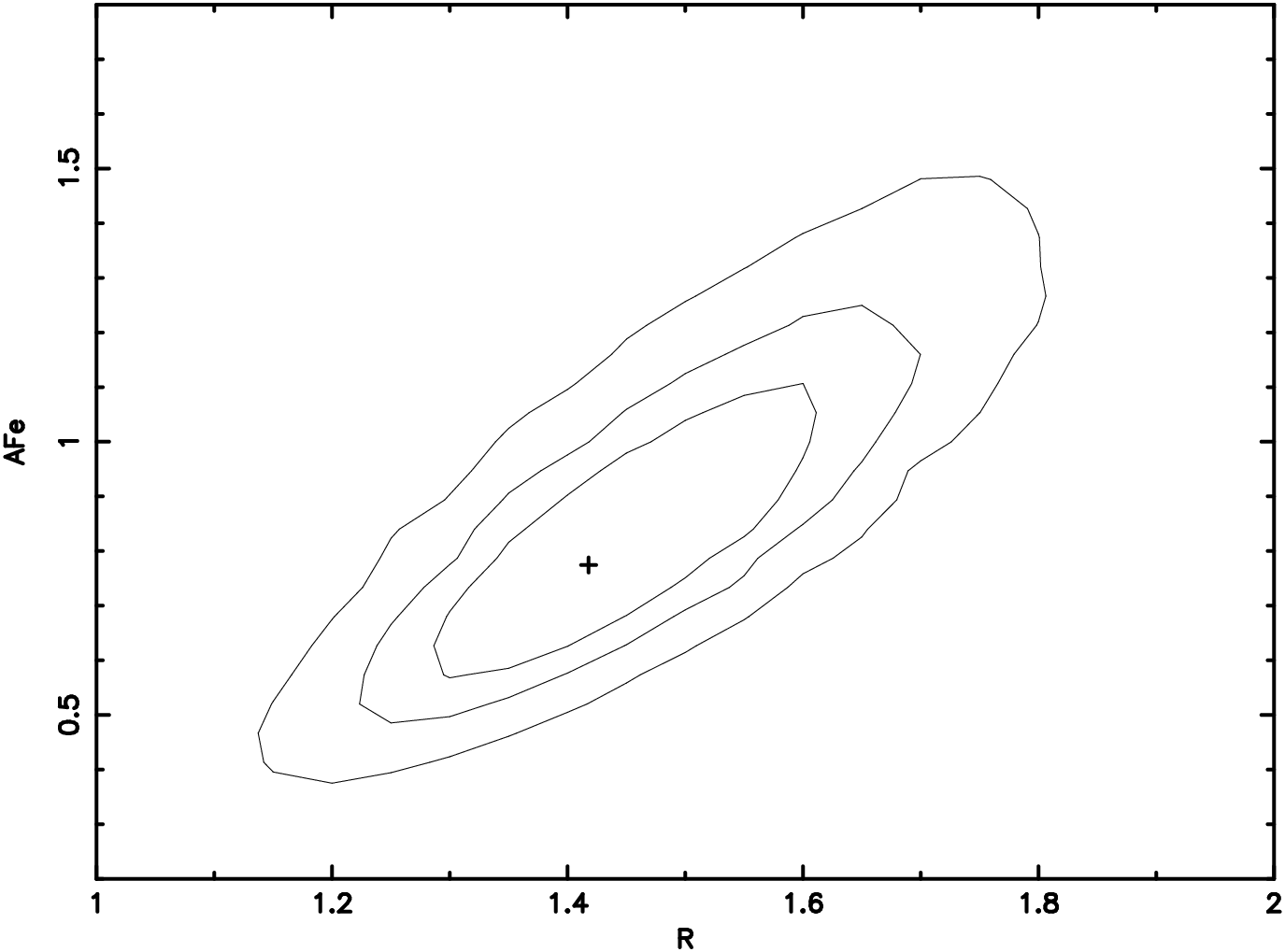,height=8.0cm,width=6.0cm,angle=-90}
\caption{$R$ and iron abundance contour plot.}
\label{r_feabund}
\end{figure}

The first is that iron is underabundant. We therefore left the
iron abundance (in units of the solar value) in the {\sc pexrav} model
free to vary. The best fit gives $A_{\rm Fe}$=0.78,
but the improvement in the fit is not significant. In Fig.~\ref{r_feabund} we show the
contour plot of $R$ and $A_{\rm Fe}$. A solution with $R\simeq$1.2 and $A_{\rm Fe}\simeq$0.5 
is acceptable
on both statistical and physical (see Matt et al. 1997) ground.

Alternatively, it is concevaible that there are two distinct reflecting regions, one
cold, distant and associated with the observed iron line, the other mildly ionized
(and likely arising from the accretion disc). 
Line emission from the latter regions would then be 
prevented by resonant trapping (e.g. Matt et al. 1996).
To test this hypothesis, we added a mildly ionized reflector with relativistic
blurring (model {\sc refsch}),
with ionization parameter fixed to 400 erg cm s$^{-1}$ (as appropriate to dump line
emission), inner and outer radii fixed to 6 and 1000 gravitational radii, and the
inclination angle to 30$^{\circ}$. 
The amount of cold reflection has been kept fixed to $R=0.7$, as appropriate given
the observed iron line EW. The fit is acceptable ($\chi^2$/d.o.f=285.9/251), but significantly
worse than that with a single cold reflector.  The best fit value for $R$, 0.25$^{+0.04}_{-0.02}$, 
is much lower than expected in a standard disc--corona scenario (i.e. $R$=1). We then allowed
the inner radius of the disc to vary, but no better fit is found.

In summary, iron underabundance seems to be preferred both on statistical and physical
ground.

\subsubsection{The Warm Absorber}

A single zone warm absorber (modelled with {\sc absori} in XSPEC)
gives an acceptable fit, but some residuals remains, 
which are well cured by the addition of a O {\sc vii} absorption edge. 
This suggests the presence of two different absorbing regions (but a
fit with two {\sc absori} models provides a slightly worse fit), or of a complex absorber. 
We also checked
that the residuals are not due to keeping the temperature of the absorber fixed to
1 million degrees. In fact, 
leaving the temperature free to vary, only a slight improvement in the fit is found.
The fit is, in any case, worse ($\chi^2$/d.o.f=249.7/252) than that with the temperature
fixed and the absorption edge. 

We also searched for a possible inflow or outflow of the warm absorber. The result is
quite disappointing: the upper limit to the outflow velocity is about 3$\times10^4$ km/s.

Of course, it is also possible that there is only one absorbing region, but that
the model we adopted is too simple. The absorbing region may well be stratified,
resulting in a gradient of ionization parameters (see e.g. R\'o$\dot{\rm z}$a\'nska 
et al. 2004 for a constant pressure model). A detailed modeling of the warm absorption in 
MCG+8--11--11 is beyond the scope of this paper. A longer exposure observation, 
able to fully exploit the 
energy resolution of the gratings, would be required to this aim.

\begin{figure}
\hbox{
\includegraphics*[width=6.cm,height=8.0cm,angle=-90]{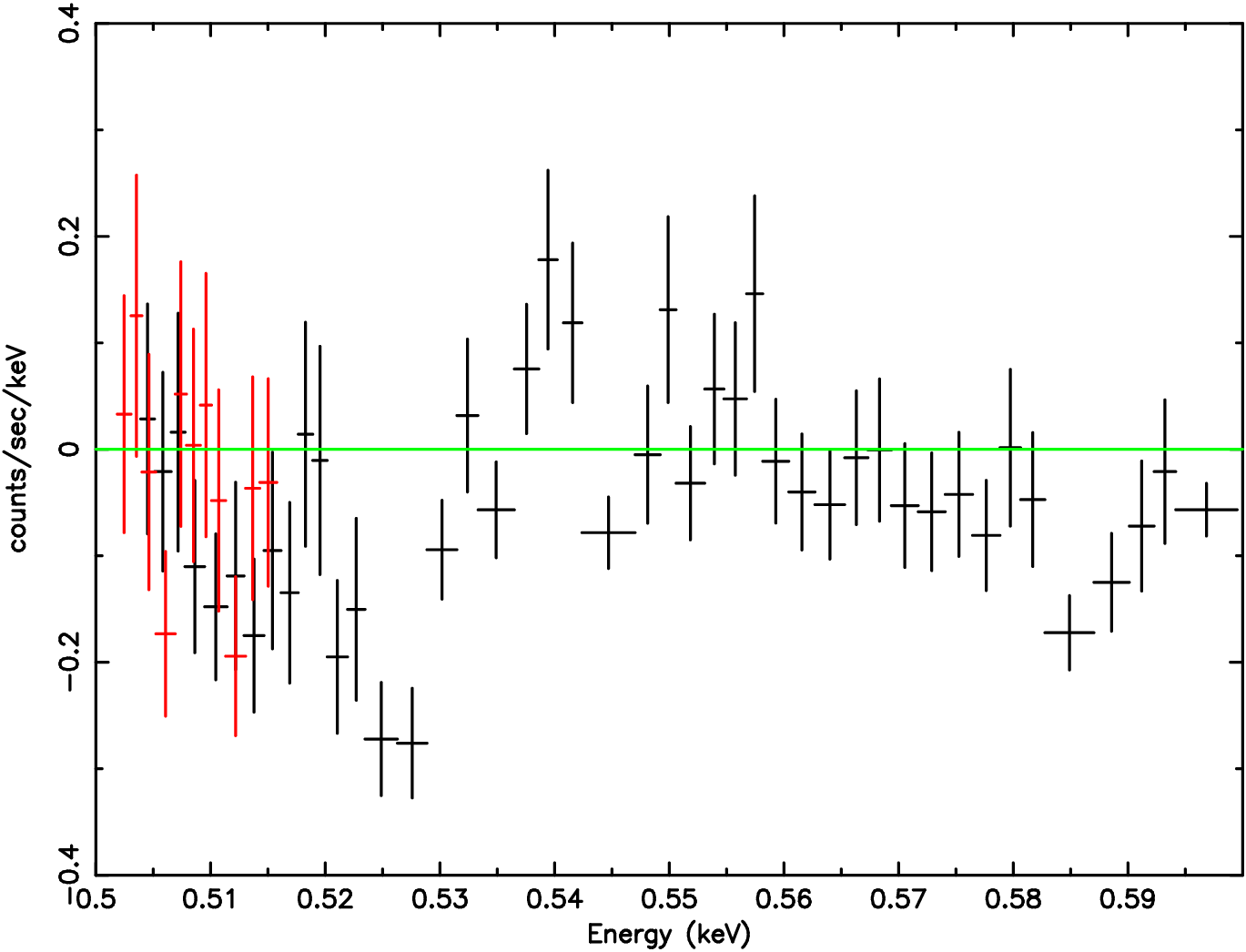}
}
\hbox{
\includegraphics*[width=6.cm,height=8.0cm,angle=-90]{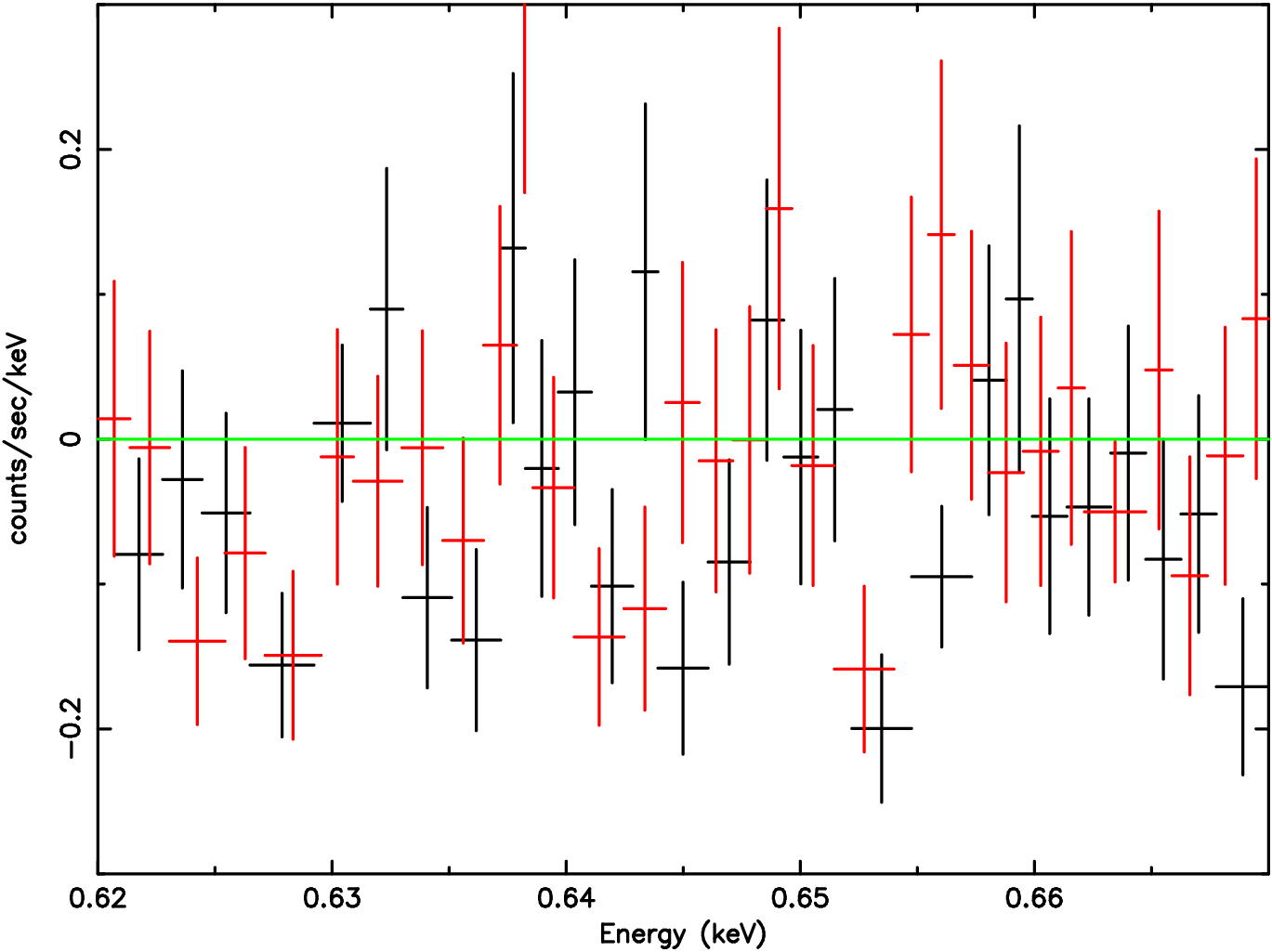}
}
\hbox{
\includegraphics*[width=6.cm,height=8.0cm,angle=-90]{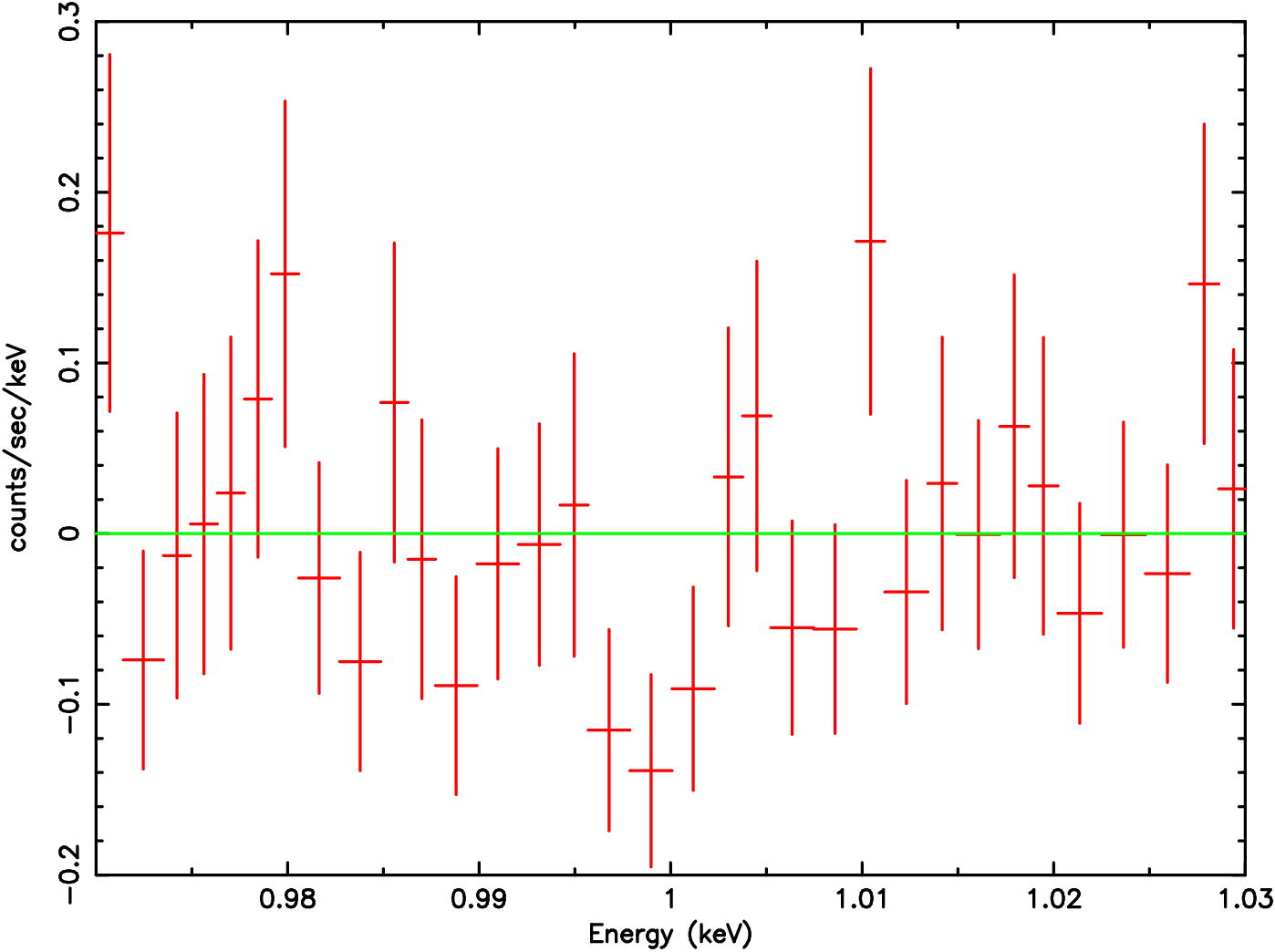}
}
\caption{ Residuals for the RGS spectra, after fitting with the EPIC/p-n best fit model, 
in the energy ranges where the absorption lines reported in Table~\ref{lines} are apparent.
}
\label{rgs}
\end{figure}

Given the brightness of the source, the RGS instruments can however still provide some 
informations on the warm matter, especially on narrow absorption or emission lines.
We fitted the RGS data with the same model used for the p-n, with all parameters
fixed to the best fit values. Inspection of the residuals (see Fig.~\ref{rgs}) 
show a number of possible absorption
lines. (We checked that these lines are still present when the continuum is fitted ``locally'',
e.g. by power laws. Note also that 
for the two of them for which both gratings were available, i.e. those at 640 and 667 eV,
the lines are visible in both instruments).
In Table~\ref{lines} we list the five lines we found with a flux percentage error less
than 50\%. The identification of these lines is not obvious. None of them are
among the possible instrumental lines discussed by Ravasio et al. (2005). 
Only the 667 eV
and the 1020 lines are at an energy corresponding to known and relatively significant transitions
(see e.g. Bianchi et al. 2005 for a list of relevant lines),
i.e. the O {\sc vii} K$\beta$ and the Ne {\sc x}  K$\alpha$
lines. The absence of the corresponding K$\alpha$ line makes,
however, the former identification quite unlikely (it must also be noted that the
{\sl observed} line energy, 0.654$\pm$0.001 keV, 
is close to that of the O {\sc viii} K$\alpha$ line, suggesting
a possible origin in our own Galaxy or the local group).
We were also unable to find a single velocity able
to provide a realistic identification even for only two lines. Moreover, identifying tentatively
the lines with the closest Oxygen lines, velocities should
be of several thousands km/s at least, 
much larger than usually observed in Warm Absorbers (e.g. Kaspi et al.
2002), even if not dissimilar to those 
observed in a few high luminosity sources (e.g. 
Hasinger et al. 2002, Chartas et al. 2002, Pounds et al. 2003a,b).  No further informations
on the velocity structure can be derived from the absorption edges, which in the RGS are
only upper limit. Clearly, a much longer observation
is needed to confirm the presence of these lines and to search for more lines and, then,
for possible systems at the same velocity.

\begin{table}
\begin{center}
\begin{tabular}{|c|c|c|c|}
\hline
~ & ~ & ~ & ~ \cr
E$_{\rm obs}$ (eV) & E$_{\rm r.f.}$ (eV) &
F (ph cm$^{-2}$ s$^{-1}$) & EW  (eV)  \cr
~ & ~ & ~ & ~ \cr
\hline
~ & ~ & ~ & ~ \cr
526$\pm$1 & 537$\pm$1 & -1.50$^{+0.39}_{-0.27}\times10^{-4}$ & -3.9  \cr
586$\pm$1 & 598$\pm$1 & -1.24$^{+0.30}_{-0.54}\times10^{-4}$ & -4.0 \cr 
627$\pm$1 & 640$\pm$1 & -0.66$^{+0.24}_{-0.22}\times10^{-4}$ & -2.4  \cr 
654$\pm$1 & 667$\pm$1 & -0.56$^{+0.21}_{-0.20}\times10^{-4}$ & -2.1  \cr 
999$\pm$1 & 1020$\pm$5 & -0.41$^{+0.21}_{-0.21}\times10^{-4}$ & -3.4  \cr 
~ & ~ & ~ & ~ \cr
\hline
\end{tabular}
\caption{Absorption lines detected in the RGS spectra with a flux percentage error
less than 50\%. Both observed and rest-frame energies are given.
}
\label{lines}
\end{center}
\end{table}

\subsection{The Ultraviolet flux}

The UV fluxes of the Optical Monitor observations were obtained by simply
converting the count rates with the method 3 described in 
the SAS documentation\footnote{\tt
http://xmm.vilspa.esa.es/sas/documentation/ watchout/uvflux.shtml}. As we are interested
in comparing the UV and X-ray fluxes, rather than in performing detailed fitting of the
Spectral Energy Distribution (SED), we did not deem it worthwhile to 
perform a more sophisticated analysis. After having checked that the fluxes for 
the three observations of each filter are consistent one another within the error, 
we obtained a mean flux of (7.33$\pm$0.04)$\times10^{-15}$ erg cm$^{-2}$ s$^{-1}$ \AA$^{-1}$
in the UVW1 filter, and of (3.52$\pm$0.06)$\times10^{-15}$ erg cm$^{-2}$ s$^{-1}$ \AA$^{-1}$
in the UVW2 filter. When corrected for extinction (after Cardelli et al. 1992), the fluxes
are (4.77$\pm$0.26)$\times10^{-14}$ erg cm$^{-2}$ s$^{-1}$ \AA$^{-1}$
in the UVW1 filter, and (4.48$\pm$0.15)$\times10^{-14}$ erg cm$^{-2}$ s$^{-1}$ \AA$^{-1}$
in the UVW2 filter.
These fluxes are roughly in agreement with those found by Treves et al.
(1990) with IUE on November 1983 (see their Fig.~8), 
in a quasi-simultaneous observations with an EXOSAT one, when the source
had a X-ray flux very similar to the XMM-$Newton$ one.

\begin{figure}
\epsfig{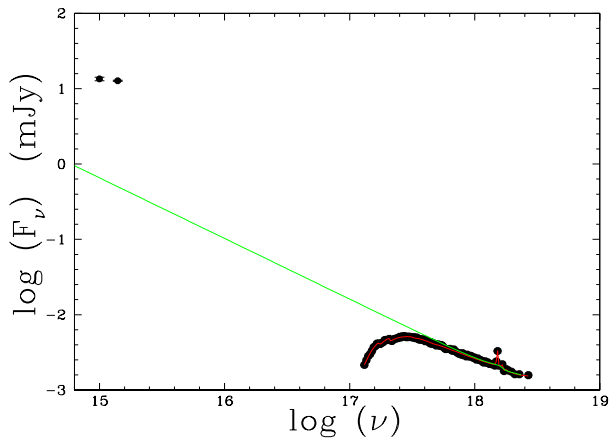}
\caption{UV fluxes as measured by the OM, along with the extrapolation of the
best fit X-ray spectrum.}
\label{sed_mJy}
\end{figure}

The UV fluxes are well above the extrapolation of the X-ray spectrum (see Fig.~\ref{sed_mJy}), 
indicating the clear presence of the UV bump. It may be tempting to take the lack of a
soft excess along with the presence of the UV bump as an argument in favour of models
of soft excesses which do not relate them with the accretion disc (e.g. Gierlinski \& Done
2004). It must be noted, however, that the large Galactic absorption makes the search for
a soft excess in MCG+8-11-11 more difficult than usual.

\section{Discussion}

When compared to the October 1997 BeppoSAX observation (Perola et al. 2000), 
the flux of the source
during the XMM--$Newton$ observation (April 2004) is about 20\% lower. The two spectra
are, however, very similar: not only the same model can be successfully applied, 
but the spectral parameters (apart normalizations) 
are also the same within the errors. The major improvement provided
by the XMM-$Newton$ observation is that
the width of the line is now much better constrained, ruling out any significant
contribution from a relativistically broadened line. 

The source was also observed by ASCA twice (Grandi et al. 1998), three days apart 
on September 1995. There was also a coordinated OSSE/CGRO  observation.
The flux was about a factor 2 lower, but again the best fit parameters are
consistent within the error with the ones we have derived from the XMM--$Newton$ data.

Therefore, despite rather large flux variations, there is no
evidence of variations of the spectral parameters. In particular, the iron line flux
is consistent with having remained constant, suggesting an origin in distant matter
(the value of $R$ which, in the same hypothesis, should have been halved in the XMM-$Newton$
observation, is so poorly determined by ASCA that no robust conclusions could be reached
on this point). 

MCG+8-11-11 is therefore 
another bright Seyfert 1 in which the presence of a strong relativistic line can
be excluded with high confidence (see e.g. Matt et al. 2001; Pounds et al. 2003c;
Bianchi et al 2004 for other cases). The lack of such lines in many, if not most, local
Seyfert galaxies is one of the more puzzling results of XMM--$Newton$. 
One possible explanation is ionization of the matter. For MCG+8-11-11, as discussed in Sec.3.3,
this solution seems unlikely. 

Alternatively, the disc may be truncated, in analogy with
the Galactic Black Hole systems in hard state (e.g. Fender et al. 2004), even if 
such analogy seems not to hold as far as Power Spectral Density 
is concerned (Uttley \& McHardy 2005). 
The lack of any measurable soft excess is certainly more easily explainable 
in the latter than in the former hypothesis. On the other hand, the UV excess suggests
that a significant thermal disc emission is present. The flux density in the two
filters slightly diminishes with frequency, suggesting we are in the region where the 
exponential rollover starts, i.e. $h\nu \sim kT_{in}$, where $T_{in}$ is the temperature
of the inner radius of the disc. Assuming the Black Hole mass given by Bian \& Zhao (2003),
a $r^{-{3 \over 4}}$ dependence of the disc temperature, and finally that the
temperature in the innermost stable orbit in a Schwarzschild disc scales with the
Black Hole mass as $T(6 r_g) \sim 2\times 10^7 (M/M_{\odot})^{-{1 \over 4}}$ K, we find
a inner disc radius of about 75$r_g$. Even if this number is different than 
the lower limit set by the iron line profile (e.g. 300$r_g$), given the large
uncertainties in this very crude procedure we can conclude that the UV flux is at least not
inconsistent with the hypothesis of a truncated disc.

\section{Summary}

We have analized the XMM--$Newton$ observation of the bright Seyfert 1 galaxy, MCG+8-11-11.
The main results can be summarized as follows:

\begin{itemize}

\item{The X-ray 
spectrum is well fitted by a $\Gamma$=1.8 power law, plus a Compton reflection component
and a Fe K$\alpha$ line. The spectrum is absorbed by warm material.}

\item{The iron line is unresolved, and likely originates in distant matter.
The comparison of the line EW and the amount of the Compton reflection component 
suggests iron underabundance.}

\item{The Warm Absorber seems to be more complex than a single zone. Absorption lines
are apparent in the RGS spectra, but their identification is ambiguous and, in any case, require
large matter velocities.}

\item{No soft excess is required by the data. The large Galactic column density towards
the source made, however, the search for such a component more difficult than usual.}

\item{The UV fluxes measured by the OM are well
above the extrapolation of the X-ray spectrum, indicating the presence of a UV bump.}

\end{itemize}

\section*{Acknowledgements}

GM and GCP acknowledge\ financial support from MIUR under grant {\sc prin-03-02-23}.
This paper is based on observations obtained with XMM-Newton, an ESA science
mission with instruments and contributions directly
funded by ESA Member States and the USA (NASA).


\begin{thebibliography}{}

\bibitem[]{} Anders E., Grevesse N., 1989, Geo. Cosm. Acta, 53, 197

\bibitem[]{} Antonucci R.R.J., 1993, ARA\&A, 31, 473

\bibitem[]{} Balestra I., Bianchi S., Matt G., 2004, A\&A, 415, 437

\bibitem[]{} Bian W., Zhao Y., 2003, MNRAS, 343, 164

\bibitem[]{} Bianchi S., Matt G., Balestra I., Guainazzi M., Perola G.C., 2004, A\&A, 422, 65

\bibitem[]{} Bianchi S., Miniutti G., Fabian A.C., Iwasawa K., 2005, MNRAS, 360, 380

\bibitem[]{} Cardelli J.A., Sembach K.R., Mathis J.S., 1992, AJ, 104, 1916

\bibitem[]{} Chartas G., Brandt W.N., Gallagher S.C., Garmire G.P., 2002, ApJ, 579, 169

\bibitem[]{} Dadina M., Cappi M., Malaguti G., Ponti G., De Rosa A., 2005, A\&A,
 in press (astro-ph/0506697)

\bibitem[]{} Ehle~M., Breitfellner~M., Gonzalez-Riestra~R., et~al., 2005, {\it XMM-Newton 
Users' Handbook}, {\tt http://xmm.vilspa.esa.es/external/xmm\_user\_support/
documentation/uhb\_2.1/ }

\bibitem[]{} Elvis M., Lockman F.J., Wilkes B.J., 1989, AJ, 97, 777

\bibitem[]{} Fabian A.C., Vaughan S., Nandra K., et al., 2002, MNRAS, 335, L1

\bibitem[]{} Fender R.P., Belloni T., Gallo E., 2004, MNRAS, 355, 1105

\bibitem[]{} George I.M., Fabian A.C., 1991, MNRAS, 249, 352

\bibitem[]{} Grandi P., Haardt F., Ghisellini G., et al., 1998, ApJ, 498, 220

\bibitem[]{} Hasinger G., Schartel N., Komossa, S., 2002, ApJ, 573, L77

\bibitem[]{} Kaspi S., Smith P.S, Netzer H., et al., 2000, ApJ, 533, 631

\bibitem[]{} Kaspi S., Brandt W.N., George I.M., et al., 2002, ApJ, 574, 643

\bibitem[]{} Lubinski P., Zdziarski A.A, 2001, MNRAS, 323, L37

\bibitem[]{} Marconi A., Risaliti G., Gilli R., 2004, MNRAS, 351, 169

\bibitem[]{} Matt G., Perola G. C., Piro L., 1991, A\&A, 247, 25

\bibitem[]{} Matt G., Fabian A.C., Ross R.R., 1996, MNRAS, 278, 1111

\bibitem[]{} Matt G., Fabian A.C., Reynolds C.S., 1997, MNRAS, 289, 175

\bibitem[]{} Matt G., Guainazzi M., Perola G.C., et al., 2001, A\&A, 377, L31

\bibitem[]{} Matt G., 2002, MNRAS, 337, 147

\bibitem[]{} Matt G., Guainazzi M., Maiolino R., 2003, MNRAS, 342, 422

\bibitem[]{} Matt G., Porquet D., Bianchi S., et al., 2005, A\&A, 435, 857

\bibitem[]{} Molendi S., Bianchi S.,  Matt G., 2003, MNRAS, 343, L1

\bibitem[]{} Pech\'a\v{c}ek T., Dov\v{c}iak M., Karas V., Matt G., 2005, A\&A, in press
(astro-ph/0507196)

\bibitem[]{} Perola G.C., Matt G., Fiore F., et al., 2000, A\&A, 358, 117

\bibitem[]{} Piconcelli E., Jimenez-Bail{\' o}n E., Guainazzi M., et al., 
2004, MNRAS, 351, 161

\bibitem[]{} Pounds K.A., Reeves J.N., King A.R., et al., 2003a, MNRAS, 345, 705

\bibitem[]{} Pounds K.A., King A.R., Page K.L., O'Brien P.T., 2003b, MNRAS, 346, 1025

\bibitem[]{} Pounds K.A.,  Reeves J.N., Page K.L., et al., 2003c, MNRAS, 341, 953

\bibitem[]{} Ravasio M., Tagliaferri G., Pollock A.M.T., Ghisellini G., Tavecchio F., 2005,
A\&A, 438, 481

\bibitem[]{} R\'o$\dot{\rm z}$a\'nska A., Czerny B., Siemiginowska A., Dumont A.-M., Kawaguchi T.,
2004, ApJ, 600, 96

\bibitem[]{} Str\"uder L., Briel U., Dennerl K., et al., 2001, A\&A, 365, L18

\bibitem[]{} Treves A., Bonelli G., Chiappetti L., et al., 1990, ApJ, 359, 98

\bibitem[]{} Turner M.J.L., Abbey A., Arnaud M., et al., 2001, A\&A, 365, L27

\bibitem[]{} Turner T.J., Mushotzky R.F., Yaqoob T., et al., 2002, ApJ, 574, L123

\bibitem[]{} Uttley P., McHardy I.M., 2005, MNRAS, in press (astro-ph/0508058)

\bibitem[]{} Wilms J., Reynolds C.S., Begelman M.C., et al., 2001, MNRAS 328, L27 




\end{thebibliography}
\end{document}